# THE LEO ARCHIPELAGO: A SYSTEM OF EARTH-RINGS FOR COMMUNICATIONS, MASS-TRANSPORT TO SPACE, SOLAR POWER, AND CONTROL OF GLOBAL WARMING

Andrew Meulenberg[a] and Karthik Balaji P.S.[b,1]

[a] NAv6 Center of Excellence, Universiti Sains Malaysia, 11800 Minden, Penang, mules333@gmail.com,
[b] National Inst. Of Tech. Karnataka, Surathkal, India, karthik.balaji1988@gmail.com

Man's quest to get into space is hindered by major problems (e.g., system-development and capital costs, expense of putting mass into orbit, trapped-radiation belts, and environmental impact of a large increase in rocket launches). A multi-purpose low-earth-orbit system of rings circling the earth – the "LEO ARCHIPELAGO™" – is proposed as a means of solving or bypassing many of them. A fiber-optic ring about the earth would be an initial testing and developmental stage for the ring systems, while providing cash-flow through a LEO-based,[2] high-band-width, world-wide communication system. A Low-Earth-Orbit-based space-elevator system, "Sling-on-a-Ring™," is proposed as the crucial developmental stage of the LEO Archipelago. Being a LEO-based heavy-mass lifter, rather than earth- or GEO-based, it is much less massive and therefore less costly than other proposed space-elevators. With the advent of lower-cost, higher-mass transport to orbit, the options for further space development (e.g., space solar power, radiation dampers, sun shades, and permanent LEO habitation) are greatly expanded.

This paper provides an update of the Sling-on-a-Ring concept in terms of new materials, potential applications, and trade-offs associated with an earlier model. The impact of Colossal Carbon Tubes, CCT, a material with high tensile strength, extremely-low density, and other favorable properties and new technologies (e.g., solar-powered lasers, power beaming to near-space and earth, and thermal-control systems) on the development of associated LEO-Ring systems (e.g., "Solar-Shade Rings" and "Power Rings") is also explored. The material's effect on the timeline for the system development indicates the feasibility of near-term implementation of the system (possibly within the decade). The Sling-on-a-Ring can provide a less-expensive, environment-friendly, mode of access to space. This would pave the way (via eventual operation at >1000 tonnes per day by 2050) for large scale development of space-based technologies.

**Keywords:** Low-Earth-Orbit; space-elevator; orbital-ring; CCT; Shade-Ring; sling

---

[1] Present address: Tektronix India Pvt Ltd, 4/2 Samrah Plaza, St Marks Road, Bangalore, 560 001, India

[2] Abbreviations:

| | |
|---|---|
| LEO = Low Earth Orbit | CCT = Colossal Carbon Tubes |
| MEO = Medium Earth Orbit | CNT = Carbon Nanotubes |
| GEO = geostationary Earth Orbit | COM = center-of-mass |
| HASTOL = Hypersonic Airplane Space Tether Orbital Launch` | T/P = tether-to-payload ratio |
| SOR = Sling-on-a-Ring | SPS = solar-power system |
| HAA = high-altitude airship | SSP = space solar power |
| Maglev = magnetically-levitated | L-SPS = laser-solar-power system |
| EDT = electrodynamic tether | sf = safety factor |
| | MT = Metric Tonne (1000 kilogram) |

## INTRODUCTION

The space-elevator system described in this paper is a crucial developmental stage of a multi-purpose, low-earth-orbit-based, system of rings circling the earth – the "LEO ARCHIPELAGO." This system of circum-Terra rings has been proposed as an environmentally-sound, economically-responsible, and technically-feasible development to solve some of man's problems on earth and to provide the foundation and stepping-stones for his move into space. The initial ring will be a fiber-optic system for broadband communications [1] and to provide a test bed for, and experience with, the dynamics of large ring systems. This ring is expected to provide near-term financial return on investment and will continue to grow with population and broadband demand.

The second ring will exploit the stability of a ring system to support a sling-based mass-transport system [2] that is essential for any large-scale system planned in space. This solar-powered Sling-on-a-Ring will allow payloads to be lifted from conventional aircraft into space without needing the inefficient, expensive, and environmentally-harmful rocket-launch systems now required to move mass from the earth to orbit. The ring, besides providing the stability and a base for solar power conversion, is a massive energy-storage medium that is available at no extra cost (financial or mass-in-orbit). With the use of electrodynamic (ED) tethers, energy and momentum can be stored (by speeding up the rings) and retrieved as needed. It can be retrieved directly, in the process of lifting mass from the earth, or indirectly (via the ED tethers) in the form of electrical energy. While the energy is converted from solar power and stored when the converters are exposed to sunlight, this stored energy can be retrieved at any time and from any location on the rings without requiring long conductive cables.

With an effective mass-lifter in place, a third ring could be an extremely-large-area "shade-ring." During initial construction, this multi-purpose ring will reduce/remove the space debris from LEO and, when extended into higher-altitude or into "slant" orbits, will eliminate the trapped radiation in the Van Allen belts. (By intercepting the radiation belts above their "mirror" points near the magnetic poles, slant-orbit shade rings will absorb, scatter, and slow the higher-energy trapped electrons and protons and nearly all of the lower-altitude radiation). With only the quickly-suppressed transient radiation from solar flares, presently-unavailable near-earth orbits (i.e., MEO) become useable and even habitable. These shade rings will be enlarged decade by decade until, by the end of the century, they will be able to reduce critical portions of the earth's temperature by several degrees Celsius [3]. They will do this from tilt orbits that shade portions of the near-polar sea ice and tundra from up to 10% of the summer sunlight. Preventing, or slowing, the seasonal melting of the sea ice is a multiplier that, by keeping the regional albedo high, can increase the impact of local shading by an order of magnitude. Keeping the tundra from melting prevents the massive amounts of (frozen) methane hydrates trapped there from escaping and becoming competition for $CO_2$ as Earth's primary greenhouse gas

As 1) power needs in space increase, 2) mass-lifting capabilities exceed priority demand, and 3) space experience with new solar-conversion technologies grows, specialized "Power Rings" will be constructed and placed into sun-facing near-dusk/dawn orbits [4]. With 100% exposure to sunlight and a uniform environment, high-efficiency systems can provide laser power to mobile units operating within the LEO system and deep into the near-earth-space environs. As power-ring capability grows beyond space needs, the decision can be made to use either MEO-ring-based or Geostationary-based (or both) space-power systems to feed Earth's insatiable need for power. The feasibility and cost effectiveness of this power ring, and of the shade ring, may depend strongly on a successful mass lifter.

Section 2 of this paper deals with the impact of Colossal Carbon Tubes (CCT) on the design and operation of the Sling-on-a-Ring. Section 3 addresses two new technologies that make Space-Solar Power (SSP) more likely (solar-powered lasers, as converters and transmitters, and high-altitude tethered balloons, as receivers). Section 3.5 presents an issue often ignored or down-played in most innovative space-system concepts: thermal control. Section 4 describes the solar-shade system (based on orbital rings) and its potential impact on global warming. The orbital-ring functionality, new CCT material, and advanced-laser technologies have a major impact in these areas as well. Section 5 is a discussion of the system: advantages, limitations, and options.

## 1. “SLING-ON-A-RING”

### 2.1 Background

Over the years, several models have been proposed for a space elevator [5] that involves mechanical-energy and momentum transfer in an attempt to reduce/eliminate the use of rocket propellants. But, few of the proposed models have gone beyond the advanced-conceptualization stage as they await technical advancements (e.g., higher tensile-strength materials and reduced launch costs). Two of the more-developed models that operate in LEO are direct predecessors of the “Sling-on-a-Ring.”

One “Orbital Ring” concept was introduced in the 80’s by Paul Birch who detailed the idea of massive rings encircling the earth in Low-Earth-Orbit. The ring is basically circular, except in the region(s) where earth-stationary “skyhooks” (high-speed maglev “trains”) deflect it. Cables, called Jacob’s ladders, hang down from the skyhooks to the surface of the earth. Although this system was an improvement over GEO-based space-elevator systems, the exceedingly-large masses involved, including very strong and bulky electromagnets, still limit the model. Since the skyhooks are stationary with respect to the Earth’s surface, they have a strong earthward force that must be balanced by the mass of the rotating ring. Reducing the mass of the skyhooks and “ladders” would lower the needed ring mass and make this system more attractive. However, reducing the mass of the ring alone (via CCT) might not be beneficial.

“Rotovator systems,” of which the “HASTOL” is the most popular/developed so far, use the concept of non-synchronous spinning skyhooks rotating with hypersonic-tip speeds. The “HASTOL” system employs a “*completely-reusable, air-breathing, subsonic-to-hypersonic, dual-fuel, aircraft that transports the payload from the ground to some intermediate point in the upper mesosphere (>100 km), where it is transferred to an orbiting, spinning, space-tether system (using a grapple assembly) and, from there, injected into a higher/different orbit*” [6]. Although the rotating skyhook can be implemented with present-day/near-future materials and the system would be much lighter than that of the Birch system, the advancement in hyper-velocity aircraft technology with the required trajectory control at such high altitudes and speeds is not yet on the horizon.

A third model, the Edward’s space-elevator concept, is an Earth-stable tensile structure that extends from the Earth’s surface to far beyond GEO. These are very-attractive features; but, the fiber-strengths required to meet the system requirements only exist in the future. Furthermore, while the proposed CNT and CCT fiber capabilities provide hope for this system, the mass of just the fiber matrix between the earth and GEO is still nearly 2 orders-of-magnitude greater than that of the Sling-on-a-Ring. (This is for equal payloads between aircraft and LEO, when the new fibers are applied in both cases.) More important than the tether mass below GEO is the tether and ballast mass beyond GEO. While the tether is of the same order of length as the 40,000 km SOR orbital ring (+/- a factor of 2), it must withstand much greater tensile forces. Relative to the earlier systems above, the Edward’s space-elevator concept has to provide a “climber” to carry its payload into space. This climber is characterized as ~3 times the mass of the payload. If the cost per weight is low enough, this may not matter; nevertheless, it could be a major multiplier (and is included in the mass comparisons above). A practical limitation of the GEO elevator is that you can’t get off at LEO; you can only stop to look around. On the other, hand it does complement the Sling-on-a-Ring (if they don’t collide).

The total mass of most elevator systems must be placed into space before the first payload is lifted without rockets. The difference between LEO and GEO in this regard is significant in terms of both cost and time required (i.e., number of launches) to place the system mass in proper orbit. In most cases, a preliminary version (e.g., lower payload) would be considered to "bootstrap" the system as a means of reducing the number of rocket launches required. The SOR is somewhat different in that the limiting factor is its "occupancy" of much of the usable LEO space, rather than its mass (see section 5).

The Sling-on-a-Ring system uses the concept of high-speed, globe-circling rings, at near zero gravity, to store energy and to stabilize a sling-hub subsystem. It consists of a thin, high-tensile-strength, equatorial, circum-terra fiber, associated power and propulsion stations, and rotating-sling modules that are integrated with the ring. It utilizes a variant of the HASTOL system to transfer a payload from the earth to LEO, but *via conventional aircraft and a skyhook-on-a-sling*. Advantages of the Sling-on-a-Ring system are its very-low mass relative to Birch's rings (or to any non-Hastol variant of space elevator) and its independence of "air-breathing and hypersonic" aircraft. Nevertheless, it does rely heavily on sling materials having a high specific strength. Since a proof-of-concept is now available for fiber materials of sufficient specific strength for the single-stage Sling-on-a Ring, this space-elevator concept has moved beyond the "what-if" stage.

Fundamental to the system dynamics of the Sling-on-a-Ring is the matching of the sling-tip velocity to that of the atmosphere. Sequential images of the sling (moving clockwise in Fig. 1) show the projection of the tip fixed near the earth and the upper end fixed at, and moving with, the sling "hub" (the center of mass moving at about 7500m/s relative to the Earth's surface).[3] At the point of pickup (at 13-15 km above the Earth's surface), the tether-tip tangential velocity is equal in magnitude to the sling's center of mass (COM) velocity (moving to the right in the figure). The net result (at sling-tip perigee) is near-zero tangential velocity of the sling tip relative to the transfer aircraft or to the earth's surface. Under these circumstances, there is no standard fiery entry of objects into the atmosphere. The actual rotational velocity of the sling can be adjusted to optimize: the horizontal tip velocity and stability relative to that of the payload-delivery aircraft (e.g., from minus 100 to plus 1000 km/hr relative to the atmosphere), the pickup window (altitude and timing), and the acceleration of the payload after pickup.

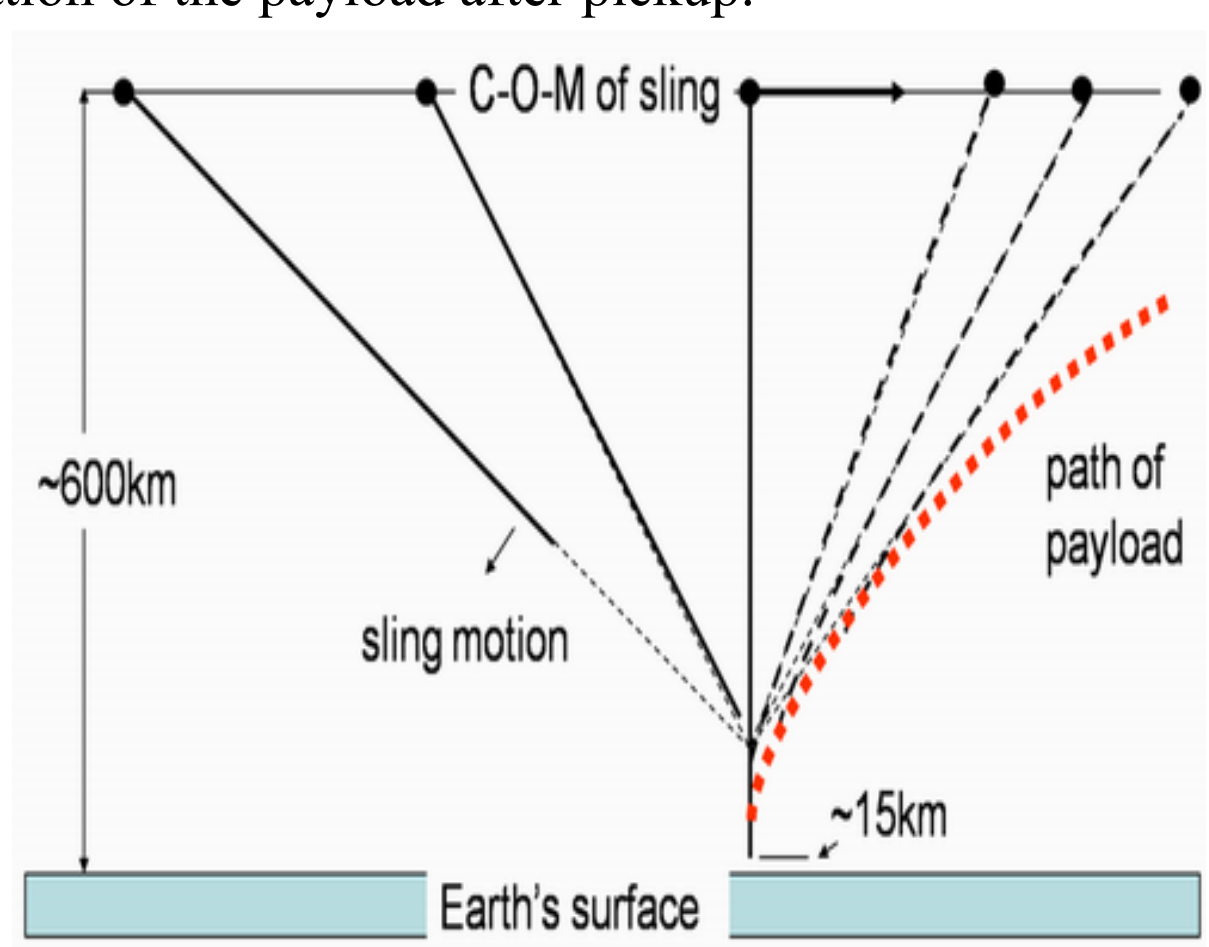

*Fig. 1 Pictorial Description of the Sling Dynamics. COM moves to the right and the sling rotates clockwise to velocity match that of an atmospheric aircraft at 13-15 km.*

The sling tip drops vertically into, and the payload is lifted nearly vertically (dashed path) out of, the atmosphere. This grappling of an object "falling" from space is a high-speed version of an art practiced in the early retrievals of high-resolution film ejected from photo-reconnaissance spacecraft. Non-computerized prop-driven aircraft intercepted and captured small packages that were parachuting into the lower atmosphere. The use of an aircraft-mounted grappling system means that the payload is not reduced by this amount. A more elaborate and secure grapple system can thus be developed without the same weight constraint as that imposed by a sling- or payload-mounted system.

---

[3] The orbital velocity (tangential, v, or rotational, ω) of the hub is slightly higher than the velocity at zero-g for that orbit {e.g. when the centrifugal and gravitational forces balance - or $mv^2r = GMm/r^2 \Rightarrow v = \omega r = \text{sqr root}(GM/r)$}.

The intercept window is less than 10 seconds long (unless various measures are taken to extend the dwell time). There is little time to maneuver. The tether is flexible, but under extremely-high tension. Therefore, if CCT fiber, it is ~ 6 millimeter in diameter at the tip (for a 10 ton payload) and therefore would be considered a thin rigid rod. Nevertheless, even with computer-controlled systems, the dynamics are challenging. However, the tip is falling vertically through a region with little atmospheric turbulence (at ~40,000 ft) and is an aerodynamic body with active control surfaces for necessary corrections. These corrections would be controlled from a computer on-board the aircraft that would be orchestrating both the tip and aircraft motions. The grapple coupling mechanism is on the payload, not on the tether. Thus, the aircraft only need intercept any point on the tether, not the tip or a specific point on the tether (otherwise the window may shrink to less than a second).

At the intercept point, the tether could still be descending (that doubles the time window). Nevertheless, the aircraft capture mechanism would allow the tether to move freely up or down for the few seconds engaged. During this brief interaction time, a counter-wrapped "cable" will be automatically applied to the tether on its "rise." Any slippage of the wrap, as the tether accelerates the payload upward (lifting it from the aircraft), will cease when the tip region (or other "stop" in the line) is encountered.

The sling is placed in an assembly integrated into a high-strength, low-specific-mass, orbital ring. The assembly (Fig. 2) roughly resembles an "H" that will serve as a platform for the deployment of the complete sling system. The system consists of a yoke, the harness "H," a hub, the counter mass (ballast), the tether, and an aerodynamic body at the tether tip. The associated solar-power and station-keeping units will be housed elsewhere on the ring for balance. A slight spin up of the ring, which puts the ring under tension (thereby serving as an energy- and momentum-storage mechanism), is required for stable deployment and operation of the system.

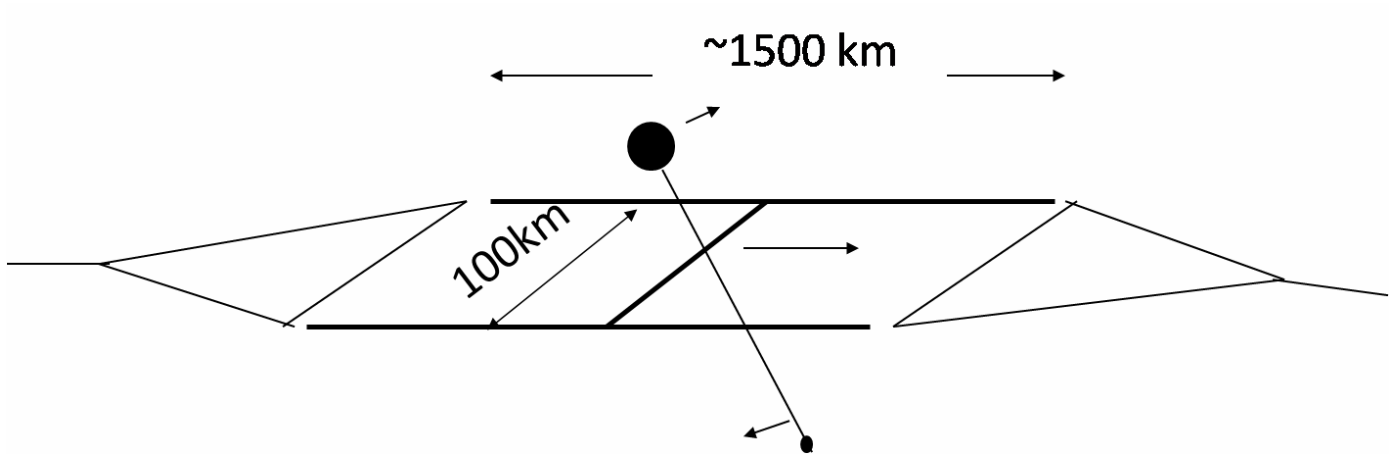

*Fig. 2 Exploded view of Sling-on-a-Ring indicating the H structure and dimensions for a 600 km long sling. The triangles join the H to the circum-terra ring.*

The hub-to-ballast connection would likely be rigid. The ballast would need to be able to move along a portion of its tether to balance angular momentum. Under rotation-induced tension, the hub-to-tether tip would also be rigid. A significant portion (e.g., 50 km) next to the hub would be structurally rigid even without rotation (to assist in spin-up and capture of the tether). After payload pickup, the C-O-M of the sling shifts toward the tip and the angular velocity of the tip tether is slowed relative to that of the ballast tether – and more so at the tip. However, the high angular momentum of the sling system and payload must be captured (mechanically) at this time for reuse in relaunching the tether for the next pickup. This momentum capture takes place through the harness and its connection with the hub. During this process, the ballast, being rigidly connected to the hub slows more than does the payload. When the payload catches up with the ballast that is equipped with a yoke to capture it, the payload tether folds about the ballast. This process continues until the payload is "reeled" in to either the ballast or to the hub (depending on the final destination of the payload).

After removal of the payload, the tip is ejected (by rail gun?) to restart the sling motion. The stored energy is applied and, in addition, the sling operating as an electrodynamic tether becomes "self propelling" with the application of sufficient voltage.

The tensile strength and the density of material making up the sling are primary parameters in the analysis of the system dynamics. The specific strength (tensile strength divided by the density of material) determines the tether/payload mass ratio [2]. For a single-stage Sling-on-a-Ring system, a 2000 ton tether is required to lift a 1 ton payload with the best fiber commercially available at that time (SPECTRA-2000, in 2008). Implementation of the single-stage Sling-on-a-Ring system evidently had problems with high tether mass although it was bypassing the need for the hypersonic aircraft required for HASTOL. It was predicted that with improvement in material strength, there would be a significant reduction in the tether mass.

In order to improve the payload-to-tether mass ratios with available material, a two-stage HASTOL system had been proposed earlier [6]. This 2-stage system would consist of a long, rotating, upper stage, through which a second tether would rotate. The 2-stage system would utilize the cross-piece in Fig.2 as its first stage. This cross-piece would be lengthened and would swing (as a deep "V") through the "H" supporting its payload, the hub (H-1) of the second stage, at its midpoint (Fig.3). The twin supports of the 1st stage would be balanced by ballast on linear structures (extending the arms of the V) beyond the H. Both the 1st and 2nd stages would be rotating in the same direction. In the Sling-on-a-Ring version, the two stages would together provide a tip velocity of ~7.5 km/s (near to the earth's surface rotational velocity) at the point of payload pickup in the atmosphere. This two-stage system became the Sling-on-a-Ring baseline because it provided a system that could operate with the materials available at the time (2008) and could grow gracefully with the improvements expected.

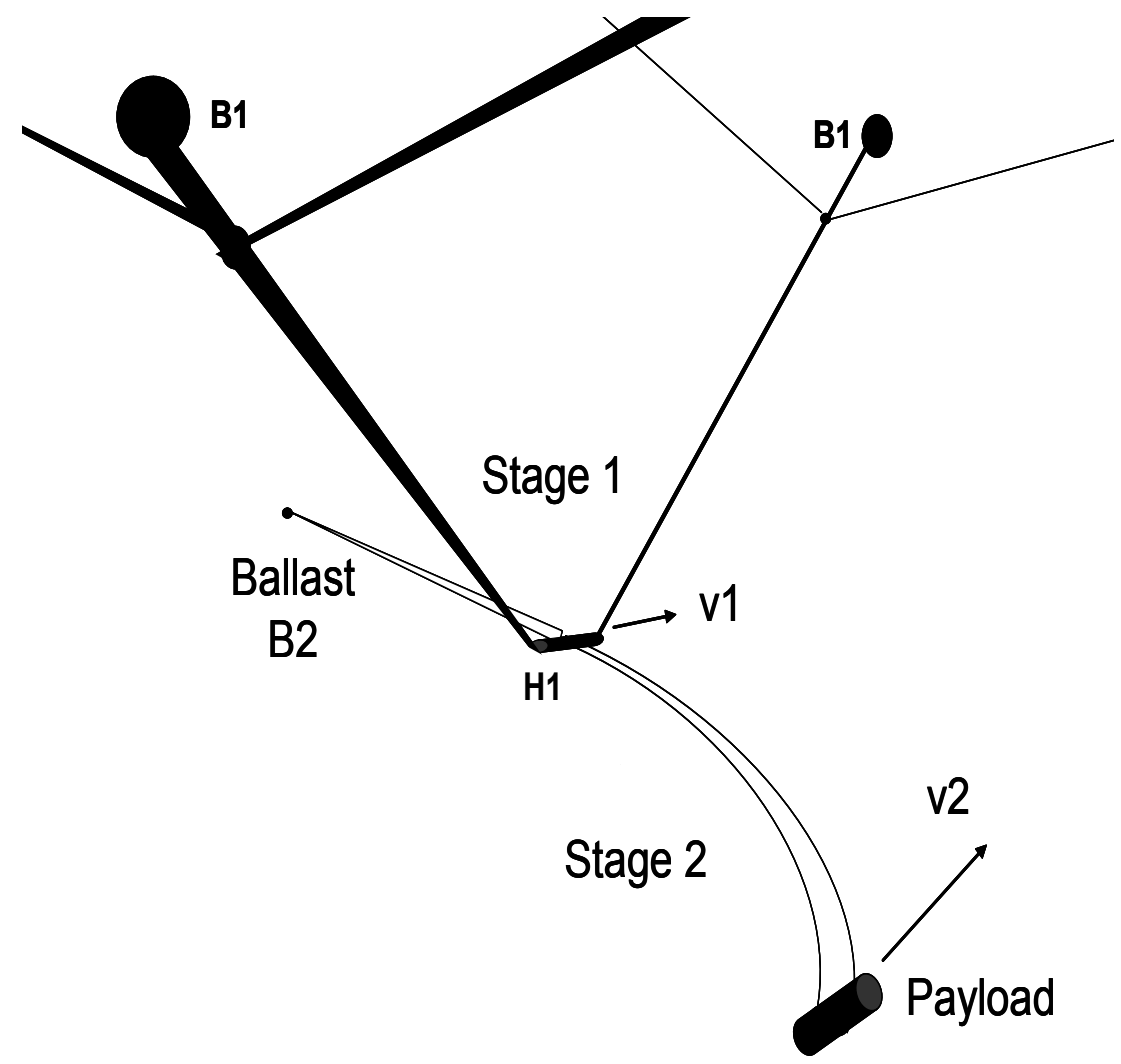

*Fig. 3 Two-stage sling-on-a-ring after payload pickup (perspective view as seen from opposite side of Fig.2).*

Most of the Sling-Ring structural components are in tension (and therefore would benefit from the new CCT-fiber technology). There are only a few components that must survive compressive forces and therefore might not be affected by the new materials. These large and more-massive components are the briefly discussed in section 5.1.

## 2.2 The Timeline and New Materials

With the advent of stronger materials, it was shown that there will be an exponential reduction in the tether/payload mass ratio. The 2008-based Timeline [2] predicted that with the expected growth in materials technology [7], one could determine a path to the super-high-tensile-strength materials within a decade. It was expected that useful tether-payload mass ratios could be achieved in space (for a two-stage system) within 10 to 15 years from beginning of the project upon validation of such material. This was assuming mass production of Carbon Nanotube (CNT) fibers with tensile strength of around 62 Gpa and density of 1.34 g/cc. But, with the validation of a new high-specific-strength material, Colossal Carbon Tubes (CCT) [8], there is a significant shift in the timeline. In the new timeline, CNT is no longer needed. (Hyper-velocity aircraft were already made unnecessary by the 2-stage Sling-on-a-Ring development). It is now expected that useful tether-payload mass ratios can be achieved in space (for a single-stage SOR system) within 10 years from beginning of the project.

CCTs are double, or multiple, -layered thin sheets of graphene that are connected by mono-layered structures to give a thin, reinforced, hollow-core compound layer that is configured as if rolled into multi-micron-diameter tubes. As such, they form structures that are small enough to exclude from the interior of the tubes any binder used to connect them. Since the tube walls are nanoscopic, and the interior voids are microscopic, most of the volume of a CCT fiber is near vacuum or low-pressure gas (as in an aerogel) from the fabrication process. Thus, relative to the presently short, small-diameter tubes of CNT, CCTs are long and nearly massless. A partial comparison can be made of the two, in terms of a long skinny balloon; CCT is the balloon when inflated, CNT when it is not. The comparison fails in that the CCT can be stretched more than the CNT.

Furthermore, in terms of a composite, with the large CCT surface areas and small relative-contact area, the amount of binder required, compared to that of the nanoscopic CNT structures, is small. (It can be made to "wick" to the fiber contact points, leaving much of the interspaces devoid of binder.) Consequently, fibers made from CCT have been observed to possess superior properties, namely, extremely high specific strength (low density – because its volume is mostly gas), excellent ductility (from the tube-dominated, rather than binder-dominated structure), and high electrical conductivity (from the constituent graphene). It is anticipated, that by control of the CCT diameters, fiber size, and film thicknesses of the material, macroscopic-structure properties can be controlled over a significant range (perhaps at the expense of fiber density).

More specifically, recent (2009) CNT macrofibers (generated by assembling individual carbon nanotubes) still show a relatively-low tensile strength in the low GPa range, while CCT macrofibers, at <1/10 the density, have already demonstrated a tensile strength of 6.9 GPa [9]. This confirms the much improved CCT mechanical properties compared to fibers of similar sizes composed of carbon nanotubes. Furthermore, a development time frame is suggested that indicates potential mass-production quantities within the decade.

Owing to its architecture, an important feature of the CCT is its ductility. Excellent ductility is very encouraging in situations requiring high toughness (a very apt material property for the tether). The CCT deforms under the tensile force in a ductile manner and the diameter undergoes a continuous local shrinkage before breaking, much like the deformation of a ductile metal wire or rubber band. This is in sharp contrast to the typical brittle-fracture nature of many advanced fibers. In fact, the CCT can sustain about 3% strain before failure, thereby giving warning before complete failure.

The combination of high strength and low density makes CCT a high-specific-strength, high-tenacity, solution to the Sling-on-a-Ring (or any space-elevator) problem. The specific strength is about 15 times that of the strongest carbon fiber (T1000). Its demonstrated capabilities now allow us to simplify the system architecture [10].

2.3 CCT vs. CNT – Single Stage system

The low density of CCT greatly increases the specific strength of the material. This has a significant impact on the tether/payload mass ratio for the single-stage sling [2]. Carbon Nanotube fibers have a density about ten times that of CCT fibers; but, they are (at some time in the future) expected to be 10 times stronger than the CCT. This section will analyze the effect of CCT on the single-stage system and consequent changes in the timeline.

The extremely-high theoretical tensile strength of CNT (~62 GPa) is compensated in the relative-tether-mass computation by the low density of the CCT (with 6.9 GPa), thus the two materials produce similar results. However, the CCT is about twice as good as the CNT (i.e., tether-to-payload mass ratio is one-half that of the CNT for tip velocities >7000 m/sec) as a result of the higher specific strength of the CCT fiber. This high specific strength of both the CCT and CNT fibers allows an increase in the safety factor (sf) for the calculation. The major hazard to the long tethers is micrometeroids. As a response to this threat, the safety factor of the tether can be increased from 2 at the hub to 4 in the thin tip region and still have a useful T/P ratio [10].

Table 1 shows the tether/payload mass ratios for the 600 km single-stage system for different high-tensile-strength materials and assuming a sling tip velocity of 7000 m/s (aircraft velocity makes up the difference to 7500 m/s). The safety factor is now graded. These T/P values are critical to the projected timeline for sling operational deployment. The low density of CCT more than compensates for the high tensile strength of Carbon Nanotubes.

| Material | Tether/ Payload Mass | Radius at tip (mm) | Radius at hub (mm) |
|---|---|---|---|
| CNT *(σ=20Gpa, sf=4)* | 100/1 | 1.8 | 110 |
| CNT *(σ=62Gpa, sf=4)* | 5/1 | 1 | 3.8 |
| CCT *(σ=6.9Gpa,sf=4)* | 3/1 | 3.1 | 8.5 |

*Table 1. Tether-to-Payload Mass ratios for a single-stage sling (sling-tip velocity of 7 km/s).*

The reduction of tether mass to "3 tonnes" for each tonne of payload, with validated CCT material, from a tether-to-mass ratio of ~100/1, with unvalidated values for CNT fibers, within the span of a year, clearly indicates a sharp change in our new timeline and also in our perspective of the system. The exponential reduction in the tether/payload mass ratio, with increasing specific strength, provides for the realization of a single-stage system without having to develop a more-complicated (and thus less reliable) two-stage system.

The corresponding reduction in the surface area of the tether for both materials, coupled with an exponentially decreasing atmospheric density with altitude, reduces the effect of atmospheric drag on the sling. However, the payload will experience some post-pickup atmospheric drag as a function of its surface area. The presence of a 100 kg aerodynamic body at the end of the line will increase the surface area at the toe end and this lower end of the sling will experience atmospheric drag along its rendezvous path. However, because of the high vertical velocity, the aerodynamic body can move the tip laterally. Because of this drag (or boost), the sling can bow as it approaches the payload. This can help to increase the available pickup time and also buffer the payload-acceleration profile.

### 2.4 Single-stage versus Two-stage system

The two-stage system was proposed [2] in order to counter the effects of excessive tether mass required to accommodate the material constraints and the very high tip velocity of an atmospheric pickup. While much more complicated than the single-stage system (in both design and operation), validation of the two-stage system parameters did show the feasibility of starting development of a testable sling project with materials available in 2008. It was shown that, by optimization of the sling lengths and by differential splitting of the velocities of the individual stages, a significant reduction in the tether/payload mass ratios could be achieved. This option allowed man the opportunity to immediately start development with the expectation of testing a "live" system within a decade, even if predicted materials and fiber-lengths did not become available in time.

The present tether-to-payload mass ratios, obtained for a single-stage system incorporating CCT, is found to be considerably less than those for a two-stage system using Spectra-2000. This one material development, which had not been expected for several years, greatly reduces the complexity in dynamics involved with a two-stage system. However, there are other implications of the change. The value of the "g-forces" on the hubs comprising a two-stage system and the payload has been shown to be close to 5-g at the point of pickup, whereas a similar calculation for the single-stage system shows an acceleration of 10-g on the payload [2]. A reduction in system mass and complexity, provided by the single-stage sling, is compensated by a higher value of payload acceleration.

### 2.5 Material effect on system dynamics and Man-rating

The payload experiences variable acceleration after pickup; but, the overall magnitude is governed primarily by the value of tip velocity and the angular momentum of the sling and payload about the common center of mass. The payload has a considerable angular momentum with respect to the central hub that is the initial center of mass and in an inertial frame. Thus, there is very little reduction in the angular velocity of the sling-tip/payload system about its initial center of mass as a result of payload attachment.

After pickup, the system's center of the mass shifts down toward the payload. This decreases the angular velocity of the ballast about the new COM. It also introduces another torque into the system. The greater the shift, the greater the retarding torque applied by the harness through the hub. As the tension from the payload travels up through the harness, the harness itself becomes part of the sling and forces the COM back toward the hub. As the harness becomes a dominant factor, the ballast can be allowed to move out to further slow the angular velocity of the sling. The new tether material reduces the value of ballast mass required to keep the COM at the hub prior to payload pickup; therefore, the shift of the COM of the sling-payload system is observably larger for the CCT as shown in Table 2 for representative systems. This becomes a significant factor in the ability to translate sling angular momentum into "lift' for the payload and to reduce and to "smooth" the payload acceleration profile. (The ability of the CCT fibers to "stretch" is a real benefit in this regard.)

*Table 2. Shift in COM position for different materials post-pickup.*

| **Material** | **Shift in C-O-M** |
|---|---|
| CNT_20GPa | 43 m |
| CNT_62GPa | 12 km |
| CCT_6.9GPa | 22 km |

The shift in COM is critical to the initial slowing of the sling, ballast, and payload rotation (it provides a torque – from the harness – in the proper direction). The system dynamics are particularly complex in the first 12 seconds after pickup, during which time the stress in the tether is transferred to the hub, harness, and ballast elements of the sling. A first order analysis of the system shows the acceleration experienced by the payload to be of the order of 10-g, thus requiring additional systems to reduce the angular momentum of the sling (post pickup) in order to make it man-rated. Some systems, operating pre-pickup, can also be beneficial. As an example, the aerodynamic tip can both increase the pickup window and alter the acceleration profile after pickup by altering the tip path during descent prior to payload pickup.

The low tether-to-payload ratio of the CCT gives greater margin with which to control the payload acceleration profile (within limits). However, if a means is not found to lower the acceleration profile sufficiently to man-rate the sling, another option still exists. Man in space will always require water. Water containers would be an important payload (even with efficient recycling). If man is enclosed in a water bath, the internal g-forces are greatly reduced and a 10-g acceleration of a single-stage sling is no longer a limitation (except perhaps to tourism). Water containers would have to be specially designed to carry passengers. However, the extra weight of such containers would not be significant with anticipated payloads in the 1-10 ton range. In the early days, and until the aircraft "payload catch" technology has shown itself to be at least 99% reliable, man's return to earth would still be via shuttle.

Unless the sling system is exactly balanced with the payload, to translate the angular momentum to lift, there will be excess angular momentum to dissipate before the sling is stopped so that the payload can be unloaded. It would be useful if a clutch could be engaged to "clamp" the sling hub to the harness at pickup. Thus, as the sling wraps about the hub and ballast yokes, its excess angular momentum is captured by the harness and transferred to the orbital ring via large diameter "spools" that "wind in" orbital-ring fiber (i.e., shortens the legs of the H in Fig. 2). This shortening helps to compensate for the orbital ring slowing down in the process of lifting the payload. The forces involved are not large, but the lengths of line "reeled in" could be significant (hundreds of kilometers). Therefore, the energy stored could be large. By reeling in the orbit ring from the trailing side of the H, which has reduced tension as a result of slowing induced by payload pickup, the energy per kilometer of stored fiber is less than that available for restarting the sling. This results because the ring has to speed back up to store energy for the next cycle.

## 3. SOLAR POWER CONVERSION & TRANSFER

### 3.1 Background

A critical need for a successful space elevator system is energy – preferably from the sun, not the earth. Since the Sling-on-a-Ring is a "quick lift" option, there is a corresponding need for very-high-power capability. The fact that the sling is on a giant ring, with its large inertia, provides the necessary lifting power. Thus, the ring is a giant, high-efficiency, high-power-density, storage unit. Solar energy can be converted into ring velocity, for energy/momentum storage, by use of solar collectors and converters followed by conversion of the resulting electrical energy into angular velocity of the ring via electrodynamic tethers [11]. The energy can be accumulated over a period of time, stored, and then abruptly extracted during the many-second lifting of a payload.

This storage capacity is not free. To store energy in the ring, it is necessary to spin it up slightly. However, since there are masses on the ring, the ring will no longer be circular once moving faster than the orbital velocity at its nominal altitude. (Increased tension in the ring will "straighten" the ring between large mass concentrations on the ring.) To minimize this problem, masses will have to be distributed in a relatively uniform manner. The ring itself, consisting of the fiber-optic communications link and the high-tensile strength supporting rings, while low-mass-per-unit-length, is still very long – and therefore massive. Since the sling assembly and associated items are also massive, and relatively concentrated, the space solar power (SSP) collection and conversion units cannot be close by, where they would localize the total ring mass even more. The solar power system (SPS) must therefore be divided into nearly equal mass units and distributed along the ring to balance the sling(s).

The initial ring will be equatorial (or nearly so); therefore, the SSP units will be in sunlight only about 60% of the time. Nevertheless, while in the sunlight, each unit contributes its energy to the ring and the sling(s) can extract the needed momentum (power) from the ring when and wherever it is needed. Thus, even in the dark, a sling can obtain the high power that it needs directly from the ring and therefore from all of the functioning SSP units. Furthermore, there are none of the losses normally associated with power transfer over long distances. Therefore, beyond being an excellent energy-storage medium, the ring is a very efficient power-distribution system.

The use of low-density, high-tensile-strength, CCT lines are not critical for ring operation. However, they make construction and maintenance simpler since, for the general applications where high specific strength is required, the larger-diameter CCT lines would be easier to handle than denser CNT lines. On the other hand, there are different "new technologies" that <u>are</u> important to the SPS of the ring(s). Two of these (solar-laser power conversion and a method for laser transmission to ground) are addressed below.

### 3.2 Microwave vs. Laser Power

Power generated in space from solar energy, if not used locally, must be transmitted for either space or terrestrial applications. This energy can be transmitted using microwaves or lasers. Pros and cons of each of these beaming methods in various parameters are reviewed briefly.

- *Size* – The laser power-beaming system has an important advantage over its microwave counterpart size-wise as a result of its smaller transmitting and receiving antennas. The beam diameter required to carry a given amount of power varies as the wavelength of the beam (approximately). The laser-power transmitter and receiver can be several orders of magnitude smaller even for megawatt power levels. (As seen below, these high power densities can be a problem.) The reduction in size brings down launch costs for the laser system.

- *Interference* – A major issue in space solar power systems employing microwave power transmission is their potential interference with satellite communication systems, which use frequencies in the same gigahertz range (because it is well suited for microwave power transmission through the atmosphere). Lasers, avoid this problem. Because of the high powers involved in power beaming, side lobes and off-frequency energies of microwave systems are a major concern for existing communications systems.
- *Atmospheric and weather effects* – With respect to power beaming to the earth, both lasers and microwaves experience atmospheric attenuation. The effect is more severely felt for lasers due to higher scattering of the laser beam by atmospheric particles of comparable size to laser wavelengths. Rain and thunderstorms are also capable of severely attenuating the laser beams (much more than they affect microwaves).
- *Safety* – Power beam safety is mandatory for both laser and microwave beams. For a given total power delivered, the power flux density of a laser beam is many orders-of-magnitude higher than that of a microwave beam. Thus, the consequences of any loss of beam control or intrusion into the beam by people, animals, or artifacts can be much more serious for the laser than for the microwave beam [12].

5.1 <u>Microwave vs. Solar-powered Laser</u>
In power-beaming systems using microwaves, solar energy is first converted to electrical power, which is then converted into, and beamed via, microwaves. The total conversion efficiency for this process is not high. A possible alternative, with increased conversion efficiency, would be by the use of direct solar pumping to generate laser power, which is then beamed, collected, and converted to electrical energy.

Conversion of concentrated solar energy into laser-beam energy has been demonstrated at efficiency levels (>30%) [13] comparable to that for solar to electrical energy conversion via solar cells [14]. (Both efficiencies can be improved by the use of selective filters on the solar concentrator; however, solar cell efficiency starts to decrease as solar concentration increases beyond a certain level [15].) From this point, laser power has to be directed and converted to DC electrical power (assume 60% efficiency for space-to-space application and 50% for the space-to-earth system proposed below). For microwaves, the DC-to-DC conversion efficiency is > 60% for space-to-earth transmission [16] so that they would appear to have an advantage; but, it is not a great one. Other factors and device optimization for the defined mission (e.g., LEO vs. GEO) will be dominant. Laser equipment is usually smaller and lighter than microwave elements, hence is perhaps more suitable for LEO-LEO power beaming (see sections3.4.2. and 3.6). When it comes to beaming down to the earth though, laser frequencies can often suffer severe atmospheric attenuation at most locations and are unsuitable. Thus, in the past, microwaves have been the only real option for power beaming to earth. Two other differences between the systems (energy harvesting and heat dissipation) are addressed (below and in section 3.5) in terms of new technologies that could be critical in how lasers and microwaves compare for the SPS applications envisioned.

<u>3.4. Proposed L-SP (laser-solar power) system</u>
3.4.1. Overview
An alternative L-SP system is proposed herein as part of the LEO-archipelago system. This concept has the potential to overcome major roadblocks faced by any L-SPS suggested for power to earth, i.e. atmospheric attenuation and safety issues. The proposed system consists of space elements consisting of spectrally-selective solar-collectors (i.e., coated mirrors), a solar-pumped laser unit, a beam director, and thermal radiators (see section 3.5). Each such L-SPS unit will be capable of generating many megawatts (10 – 20 MW) of beam power. The earth segment would consist of a medium altitude (5-6 km) tethered "balloon," an aerostat, with a concentrator solar array to receive and efficiently convert the monochromatic laser beam to electricity (see 3.4.3). The wavelength of the laser bands would likely be selected at a strong lower-atmosphere-absorption band in the infrared region (e.g., ~1365 nm). However, water vapor absorption (with little else, other than methane, absorbing at that wavelength) would be ideal, since the aerostat would be at temperatures below the freezing point of water. Therefore, outside of clouds, there would be little water vapor above the aerostat and high absorption from that which is below. Since the laser beam is monochromatic, its interaction with specific absorption lines, its line width relative to the absorber's, and its saturation effects would all be important.

By year 2100, there could be hundreds of thousands (eventually perhaps millions – 10M x 10MW = 100 TW) of L-SPS units spread across several "power rings" as part of the LEO archipelago. The power ring(s) would be sun oriented (i.e., in a 100% sun-facing near-polar orbit – normal to the sun-Earth axis – that precesses at the rate of about 1 deg per day so that it continues to be fully illuminated as the Earth moves about the sun).

3.4.2. LEO-to-Space energy transfer

To avoid interference with the sling and communications rings, the power rings are likely to be out beyond 2000 km. The initial, and one of the most important, use of the laser-power system may be space-to-space power beaming to support the Mass-Lifter ring (the sling ring at ~600 km above the earth's surface) and any unattached (non-ring) assets and infrastructure growing in the newly available near-earth space. While the sling ring will have its own SPS, and can provide power for local needs (on-ring or off), the greater efficiency of a dedicated ring, further from the earth and exposed to sunlight 100% of the time, would mean that space tugs, free flyers, construction sites, etc. would have access to less-expensive, high power, monochromatic light for efficient (>60%) conversion to local electrical power with small receivers.

The sling ring can provide the means of injecting payloads into transfer orbits, either directly from pickup or indirectly, via separate sling or rail gun with a stopover at the sling ring. This initial input cannot put a payload into useful orbit; it provides the transfer orbit from which the payload must be moved into the final orbit. That must be done with onboard rockets, with ion-engines, or with the help of space tugs. These space tugs or space buses, needed for the new development in near-earth orbit, are best powered from the more efficient and higher-orbit power ring. The possibility of solar-lasers, whether on the power ring or the sling ring, makes a major difference in this option.

3.4.3. LEO-to-earth energy transfer and communications system

The use of tethered high-altitude aerostats (at >5 km) as power-receiving stations is proposed to enable LEO-to-earth laser-power beaming; thus bypassing most atmospheric attenuation/scattering problems. The goal is to get close to 10MW to ground per aerostat. High numbers of them permit system diversity, which avoids other problems and losses. The laser-power beams (at wavelengths coincident with one of the atmospheric water-vapor bands in the near (or far) –infrared) would be permitted only at oblique angles if attenuation of several decades is not possible with a single thickness of atmosphere. The oblique angles greatly increase the path length in the atmosphere so that no beams could hit the Earth's surface without being greatly attenuated. The proposed aerostat has an operational altitude around 5 km, where its laser reception would not be much affected by (liquid) water that would generally be frozen. But, any "spillover" of the beam would be highly attenuated in its long slant-path to the earth by the increasing concentration of water vapor with depth into the atmosphere. The aerostat tethers could have super-conductive power lines running through them to carry the resulting electrical power (e.g., 20A at 600kV) to the earth. (The cable mass vs. resistive losses and total system efficiency has implications for the whole power chain and is one of many tradeoffs required in the system studies.)

The new CCT fibers can provide greater strength and conductivity than present materials and would therefore reduce aerostat-tether mass. This would allow greater payload or higher operating altitudes. Since the laser beam is monochromatic, a single-junction, 5x4cm, monolithic structure, patterned as a micro-array of high-efficiency concentrator cells, can provide 1 kV output per unit with correspondingly reduced currents [15]. A 10 $m^2$, 10 MW-output array would contain ~ 4000 of these micro-arrays. In series and parallel connections, they would provide output voltages of 0.2 – 1 MV with corresponding currents at the 50 – 10 A level.

The aerostats would not be single purpose. On some, a small top-mounted lasercom unit (send/receive) would provide broadband connection to the communications rings (at 600 km and above 1500 km) and to the power ring (at > 2000 km). Small microwave horns (for long-range, terrestrial-cellular, communications) would be placed around the edges of other aerostats. All of the aerostats would have fiber-optic links built into their tethers. The end result is that while the primary purpose of the aerostats is to provide power locally, the multiplicity of these units (remember eventually there will be millions of them), provides a dense, highly-interconnected, communications network as well. The system would allow developing nations to move into the modern-communication age without paying the price of a wasteful hardwire-based power and broadband-communications infrastructure (just as cell phones have allowed much of the world to bypass the need for telephone lines). Nations with established infrastructure will use this option for growth and eventual replacement of their present systems.

The use of laser-power from space to fuel a high-altitude airship (HAA = non-tethered, self-powered, blimps) for fuel-efficient transport, communications, and other long-duration applications [17] is a variation on the theme described. However, since the power-handling and efficiency requirements are so much lower, the ability to use top-mounted, fixed, solar panels (no array directors required and therefore much-lower weight) and powered from GEO may make more sense. With sufficient successful experience with these pollution-free air-freight carriers, the transition to high-intensity beams from GEO to aerostats might be a preferred (or complementary) option to the use of oblique beams from a LEO power ring. With the Sling-on-a-Ring as a less-expensive mass lifter and the shade rings suppressing the trapped radiation, GEO and lower orbits (MEO) become much more accessible, and therefore more attractive, for many applications not presently available or cost effective.

### 3.5 Thermal Control & Space Power

Thermal control of space systems is a critical and difficult problem. Yet, it is seldom addressed for newly proposed space-based, mega-power systems [18]. Since few elements of an SPS can operate at high temperatures or at close to 100% efficiency, thermal control (based on radiative power loss from radiator panels to space) must be employed at each stage. Fortunately, one exception is large-area, non-concentrator, solar-cell arrays which provide sufficient cooling for operation in the 300 K range.

When cooling is required and assuming that scaling holds, the choice for transportation of waste heat (to the radiator surfaces themselves) is high-throughput heat pipes. Heat pipes are used to 'transport' the heat from a source to distant location (the radiator) to be emitted at a similar temperature. This non-conduction mode of heat transport makes large radiators feasible (and therefore solar concentrators and moderately-efficient, high-power technology). Present heat-pipe design allows thermal distribution over the full range of temperatures that must be considered.

The use of carbon composites [19] (now including CCT- and CNT-based technologies) is considered for the heat-pipe shell and thermal-radiator material. These new materials, and the possible extension of existing or new heat-pipe technology [20, 21], hold strong promise for future thermal management in space (and in the aerostats). Based on the low densities and very high values of thermal conductivity of these new materials, heat-pipe and radiator masses can be reduced significantly.

If SPS elements can operate at high temperatures, then large area radiators are not required for that element. This has an obvious impact on savings in complexity, area, and mass of the radiator. Note that, without concentration, solar cells meet the 300K operating temperature criterion, if they have infrared reflective coatings and can radiate from both sides (radiator-to-source area ratio = 2).

The possibility of solar-lasers being able to operate at 900K gives them a tremendous advantage over microwave systems that must be powered by systems that have much more stringent operating-temperature requirements (e.g. ~300K). On the other hand, the use of receivers on the earth (solar cells for laser beaming and rectenna arrays for microwaves) does not impose the same type of penalty. The 60% efficiency of the cells in a -30C forced-air environment is not a severe limitation in terms of cooling – even at the MW level. Nevertheless, in a high-altitude aerostat, weight is always a limitation and mass-efficient cooling is therefore a necessity.

A 10 MW laser-beam converter in space would require a ~3,500 $m^2$ radiator for a 70% efficient PV detector array. The ability to radiate from both sides would reduce this area by a factor of 2. But, since the radiant power source is not the sun, the radiator array does not automatically have a shade to "hide" behind (e.g., when not in the earth's shadow). Provision of low-absorption, high-emissivity, radiators is another system tradeoff that depends strongly on the affect that new materials have on the radiator mass and efficiency. This PV-receiver radiator could be large relative to a comparable-power microwave-rectenna field and <150 $m^2$ radiator required for a comparable-power, but 90% efficient, microwave beam convertor (assuming that it can operate at higher temperatures, e.g., 500K). Thus, the apparent "no-brainer" of using high-power lasers for intra-space power transfer is no longer so obvious. On the other hand, relative to laser-beam, the microwave-beam transmit and receive antennas/fields are still much larger and must also be directable. These latter sizes are strongly dependent on the required efficiencies and the distance at which the power is to be conveyed. Therefore, a major system study is needed to determine the best option(s). Table 3 provides a very- rough comparison of these systems.

Table 3. Comparison of thermal radiator and transmit/receive arrays for laser and microwave systems.

| 10MW converter | Efficiency (%) | Thermal Radiator Temperature (K) | Thermal Radiator area ($m^2$) | Transmitter Array size ($m^2$) | Receiver Array size ($m^2$) |
|---|---|---|---|---|---|
| Laser | 70 | ~300 | ~3500 | >1 | ~10 |
| Microwave | 90 | ~500 | < 150 | Large to small | $10^3$ to $10^4$ |

Particularly for large radiators, the introduction of CCT for the structural members and heat pipes and graphene in the large area thin-film radiators would have a major impact on this critical item. It can make the difference as to which power-transfer system is chosen for various missions.

New technologies, developed to meet other needs, can often be modified and applied to the specific needs of a space civilization. The high-altitude aerostat, manufactured in the tens of units for current terrestrial applications, could be required in the millions, if space-power systems become a reality. MW-capable, iodine-based lasers [22] (with emission at 1315 nm, nearly coincident with a water absorption line [23]) have promise. Cooling system designs, which have a long history in space, would need to be modified to handle scaled up radiator and heat pipe technologies. New materials and technologies can help improve them, but their lack is not a "show stopper" (unless scaled versions of present systems are not acceptable).

### 5.1 Safety issues

The aerostat/laser-power system will not be without critics. A major (and legitimate) argument against it will be one of safety. MW/$m^2$ power densities of non-visible power would be expected. Even if attenuated to 1% of its initial power, the resulting beam has an order of magnitude more power density than available on earth from the sun. One would not need to be looking directly at it, or a reflection, to be permanently blinded (and it is not visible, so there is little warning).

This is not a system that could be applied to the earth environment without a long history of near perfect operation and / or proof that the beam cannot come within a kilometer of the earth's surface. That is why its use in space-to-space power transfer over an extended period would be a prerequisite for any laser-power-to-earth applications. Its use in space would also have to prohibit transfer to any receivers with Earth as a background, if non-penetrating wavelengths cannot be provided.

Figure 4, using laser beams between the power ring and the sling-ring environment, indicates the ranges of power transfer permitted assuming penetrating beams. The left side of the figure indicates north/south power transfer from various points on the power ring to a single point on the sling ring (at 600 km). The right side indicates the east/west power transfer from various points on the power ring to various points on the sling ring. It is for this reason (limited exposure to receivers below the laser source) that could make laser power-beaming units necessary on the lower-orbit sling ring as well as on the power ring.

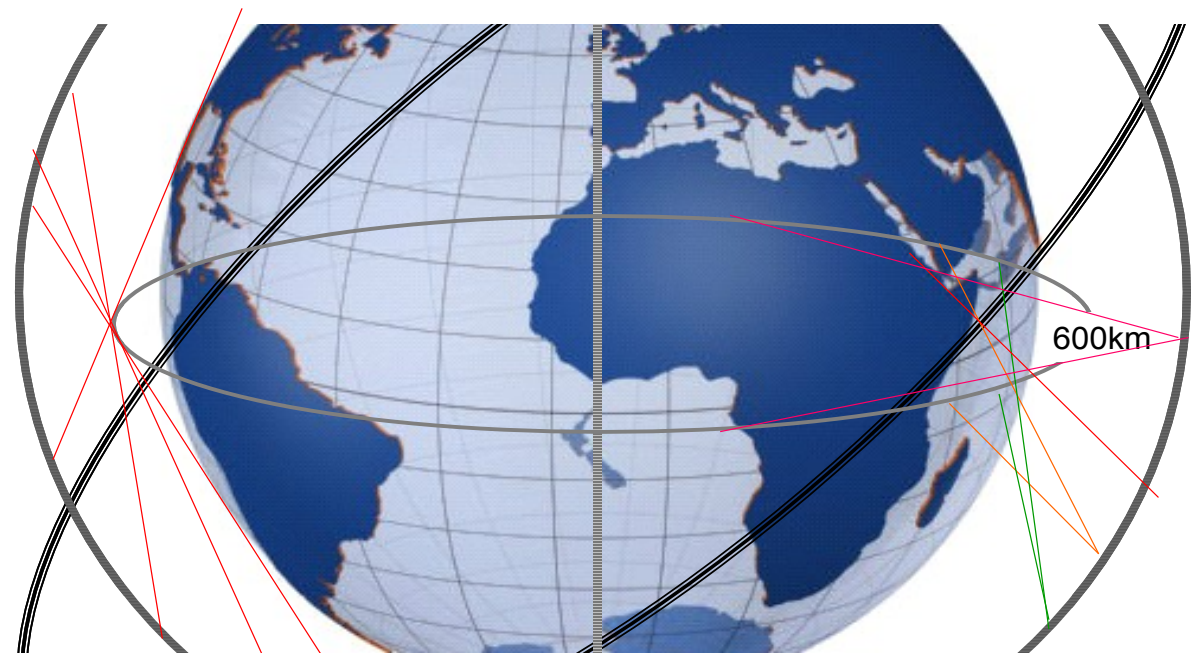

*Fig. 4 Permitted laser-beam angles between a power ring (facing out) and the equatorial sling ring. A higher-orbit shade ring (slanted) is shown also.*

Since power transfer to earth-bound receivers (aerostats) must likewise not hit the earth, the laser beams incident on the aerostats must be nearly horizontal, if they are not fully attenuated by the lower atmosphere. This is one of the reasons why the present aerostat systems will work. The photovoltaic array and director is the heaviest part of the laser-receiver payload. A side-viewing array cannot be carried as stably on top of an aerostat. On the other hand, if a GEO laser source is determined to be an acceptable option, then a simple horizontal-concentrator array (no director required to point straight up, 3.4.3) would have many benefits.

Figure 5 is a picture of a presently-available candidate for this role, TCOM's 71m aerostat [24]. The power receiver is placed in the "bulging structure underneath the aerostat". The receiver-section sides would be highly transparent and anti-reflection coated at the laser wavelength (perhaps slightly fritted to provide low-angle scattering to make illumination of the cell array more uniform) and will house the laser-beam receiving equipment. The ~4 m diameter PV solar array (3.4.1&3) is small compared to the 4x7m radar dish that is presently operational in these aerostats. Since there exists a narrow usable window of angles that are both possible and permitted for these high power beams, this system would be a safe, yet realizable, method of laser power beaming from LEO to earth.

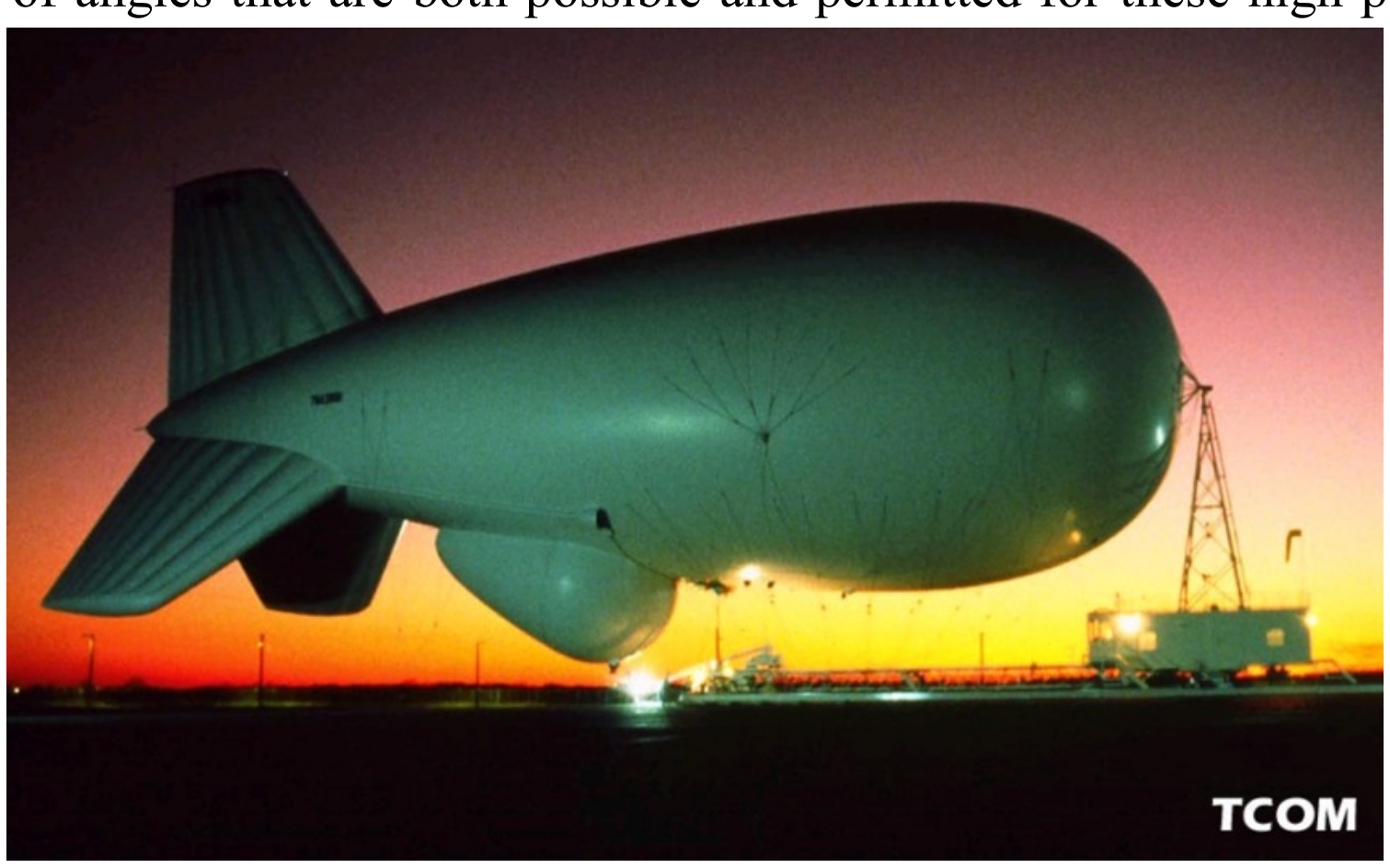

*Fig. 5 Large, high-altitude, aerostat capable of laser-beam reception and power conversion. (Photo courtesy of TCOM, L.P.)*

The hazard to aircraft of a large number of high-altitude "barrage balloons" will certainly be brought up [25]. The answer here is that randomly placed streetlights would also be a greater hazard than benefit to cars traveling at night. The aerostats will be below commercial air traffic; but airports will present a special problem. The arrangement of tethers and the associated laser beams will lead to the creation of corridors in the airways. Air-traffic control will be vital. Flight control will likely be fully computerized.

These laser beams may not “zap” airplanes, but the windows (and perhaps most surfaces) will need highly-reflective (>99%) narrow-band filters to prevent mishap if an aircraft should intercept a beam. Even at 1% absorption, a 1.5 $MW/m^2$ beam gives >10 suns intensity. In a few seconds, that level could thermally distort an aircraft or damage eyes of anyone looking out of (such coated) windows. The presence of these laser beams will thus demand the presence of a better coating on all LEO satellites and spacecraft and a highly-reflective, non-transmitting coating (in the laser spectral range) on all aircraft windows. However, feedback controllers at the laser source (sensing beam reflection from an aerostat-mounted corner cube) would shift, or cut off power beams from LEO in 10s of milliseconds.

In addition to these problems, there is always scope for misuse of such a laser-power beaming system. By changing its wavelength slightly (to an atmosphere penetrating one), it could be directed to wipe out people, or to vaporize many structures, on the earth. Thus, if such wavelength changes are possible, high security and international agreements are required, so that this power can never be abused. [The use of a chemical laser would preclude the ability to change wavelength sufficiently to present such a danger.]

## 4. SOLAR SHADES

One potential driver for the Sling-on–a-Ring is the need for controlling global warming in a socially-, economically-, and environmentally-acceptable manner. The deployment of solar-shade rings to mitigate global warming is an option. In addition to selectively-controlling thermal input to the earth, these large-area solar shades, even during their early deployment and growth in LEO, would provide benefits such as reduction of space-debris and depletion of the Van-Allen radiation belts.

The effect of the deployment of large-area, low-mass, thin sheets of opaque or reflective material (such as 4-6 micron-thick aluminized Kapton) into rings at high-LEO has been modeled [3]. This shade-ring structure would block or reflect away a small fraction of the total sunlight reaching the Earth, hence causing a temperature drop on the surface of the Earth. Key to the concept is that the location of this change in temperature can be selected to a great extent (unlike other proposed solar-shade models). Thus, even though only a small fraction of the solar energy is blocked, the consequences for global warming can be significant. The remainder of this section will give an example of such selective shading from near-polar rings on global temperature patterns using a simple model.

The Terrestrial temperature profile has been approximated for each 10° latitudinal zone with a one-dimensional model. A shade ring is proposed for an altitude of 2000 – 4000 km and could consist of thin-film mega panels totaling up to 4% of the earth’s surface area (to block ~1% of insolation [26]). Maximal total cooling, via specific orbits (e.g., polar or equatorial), is not necessarily optimal. The polar orbit (Fig. 6) gives a uniform cooling to the whole earth (summer and winter). The equatorial orbit maximally cools only the equatorial region and could have either positive or negative consequences on the world-wide environment as that region strongly affects both the weather and ocean currents. Slant orbits (Fig. 6) can provide cooling to only the summer hemisphere, but at the cost of less total shading of the Earth. Here, specific emphasis has been laid on the selectivity effect in the sub-Polar regions. The “tilting” of the solar-shades (Fig. 6), to reduce its cooling effect at the poles (where it is not needed) and to increase it in the near-polar regions (where it meets known needs with a minimum of negative consequences) is recommended. Even though the total cooling is less than that of other orbits, the benefits outweigh the lowered cooling efficiency of the shade. Shading from a ring in an appropriate tilt orbit mainly limits the effect to only a couple of the latitudinal regions (e.g., the 75° and 65° bands) where it is possible to reduce the summer melt of ice. Since the polar-most region (>80°) has more thermal “cold” stored and is further from any warming ocean currents, its ice can perhaps survive Global Warming without being shaded, particularly if the adjacent regions can be more-highly shaded as a consequence.

*Fig. 6: Three options (of many) for solar-shade rings: polar orbit (vertical); slant orbit; and tilt orbit (near horizontal). Viewed from the sun, the length of a tilt-orbit shadow (in red) cast over the critical polar regions (80$^{o}$ > Latitude > 65$^{o}$) is increased ~6x, relative to the polar-orbit shadow*

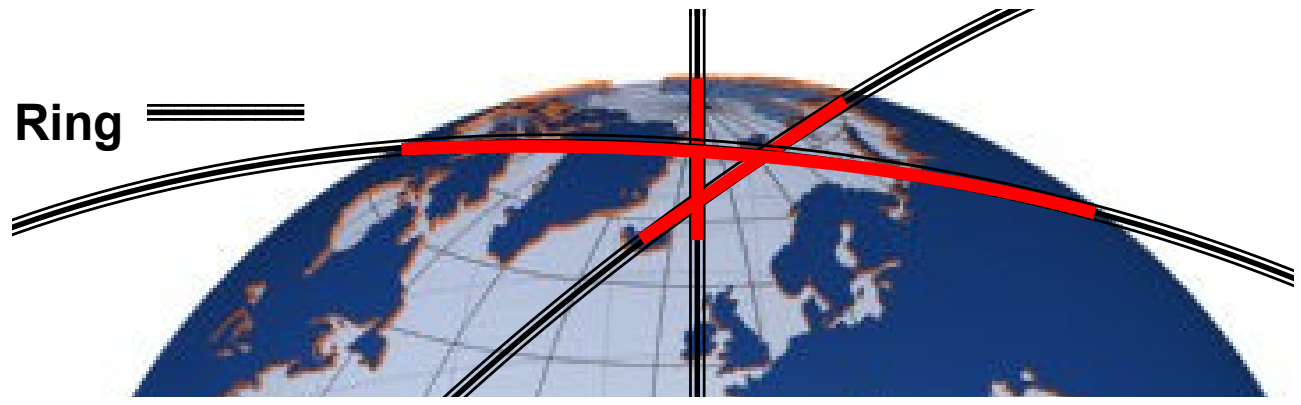

As seen in Fig. 7, one such proposed tilt-ring (triangles) could reduce the peak summer temperature of the sub-Polar regions (~75° latitude) by > 3°C. The polar ring (squares) is seen to cool the poles (both of them, summer and winter) more than does the tilt-ring; and it would cool equatorial regions more as well. However, this greater overall cooling is not as important as the selective cooling provided by a tilt-ring. If melting of the sea and tundra ice can be prevented or delayed, preservation of the high albedo of ice and snow prevents local and global warming far beyond the apparent shade-induced change in temperature.

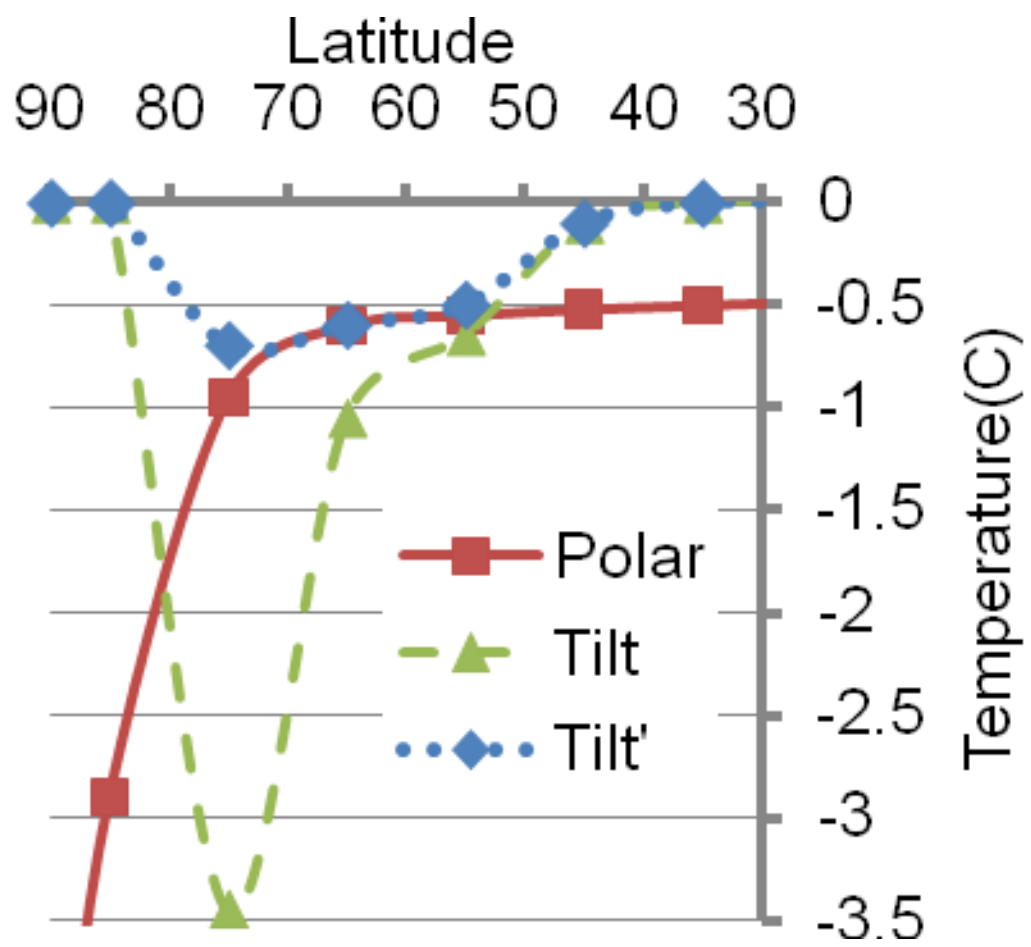

*Fig.7. Shade-ring effects (degrees C cooling as a function of latitude) for different shade-ring orientation (triangles for tilt ring, squares for polar ring) and for a different thermal mass and latent-heat model (diamonds).*

The actual change in local temperature depends strongly on the shaded-surface conditions. If they are only ice and snow, their high albedo reduces the change in temperature of the tilt orbit (Tilt', diamonds) because the low initial absorption of light means that the shade has less consequence. If the surface is water or bare land, with low albedo, the shade has the much greater effect on temperature, as seen in the triangle curve. The two tilt orbit curves block the same amount of solar radiation from hitting the earth. By changing the tilt angle slightly, the short-term effect on temperature can be strongly increased. But, by forgoing the short-term gains, it is possible to delay the melt. Thus, the shade ring prevents the long-term increase in solar-energy absorption associated with the bare sea and land. Therefore, a time sequence and integration is required to see the overall effect. A more sophisticated model is necessary to better determine the actual changes and annual impact for the various conditions and ring orientations.

The orientation of shade rings can be further refined (e.g., seasonally adjusted) depending on the shade size (e.g., during its construction) and on the terrestrial requirements. For example, prevention of a melting of tundra (in Siberia and Canada) is vital in the effort to maintain the frozen hydrated methane (a most significant greenhouse gas) in the ground or seabed. A balance between near-term and long-term goals must be maintained. The shade ring is nearly unique in its ability to do this "fine tuning" of the solar input. With ring orientation fixed in space, the annual rotation of the Earth's axis would give similar shading to the southern hemisphere (to protect the Antarctic ice sheets) as its summer approaches. In between these critical periods, the tilt orbit becomes a slant orbit and provides more total, non-specific (global), shading. Again, proper adjustment of its orientation could give preferential cooling to the summer hemisphere (at the cost of total global cooling).

With such a large shade system, it is possible to consider a "Twilight Ring." This Ring is the portion of a Shade Ring on the backside of the earth. If the sun-side coating of the Shade Ring is reflective in the visible range, then the nonuniform surface of the backside ring (until it enters the Earth's shadow) will scatter useful sunlight back to the earth in the winter hemisphere. The unshadowed portions of this large-area reflector would be much brighter than the moon. Furthermore, the "backside" of the Tilt-Ring, all of which is oriented toward the sun for maximal shading, would be focusing light toward the earth. (This is assuming that an efficient means, e.g., dynamic, is found to twist the shade on a rotating ring so that it maintains its full face to the sun.) The amount of focusing can be somewhat controlled by making the Tilt-Ring elliptical. Therefore, the short winter days would be lengthened by two hours of extended bright twilight on both morning and evening sectors. The amount of electricity saved by reduced lighting requirements would greatly exceed the reflected solar-energy deposited by the Tilt-Ring backside.

While the proposed system seems to be science fiction, or at least a far-future event, the required technology and materials have already been demonstrated.

## 5. DISCUSSION & CONCLUSIONS

### 5.1 Arguments against the LEO Archipelago - and response

Safety issues related to the Laser Power-to-Earth option have been discussed (3.6). However, there are issues that could prevent any of the ring system from ever getting off the drawing board. Financial issues can be addressed from the viewpoint of initial early returns of investment from the Communications Ring and the much-larger mid- and long-term profits from the Sling-on-a-Ring, Power Ring, and Solar-Shade Ring. Political issues of national and competitor's rights; sovereignty; and access to space and its benefits must all be addressed. Before any of these concerns are addressed, another one must be resolved. What does the Ring System do to use of LEO and to transit into higher orbits by other systems.

Even early deployment of the Communications Ring would be a hazard at its chosen altitude. However, as it grows and moves up into higher orbits, it would be out of the way for nearly all space missions with simple care in orbit control during launch and affected transfer orbits.

There are limitations to, and benefits from, all of the proposed space elevators. All of them will interfere, to a greater or lesser extent, with present usage of LEO (and some to all higher orbits). The Sling-on-a-Ring, because of the orbital ring, but more so because of the sling, may interfere more than most. If it operates at 600 km with a ~600 km sling, it has the potential of damaging anything in orbit that drops below 1200 km. Careful "traffic" control would allow some joint use of this region. Nevertheless use of LEO would be severely limited in terms of available orbits, timing, and mission length. Until the sun-shades start being put into orbit and suppress the radiation belts, the SOR would preclude most present missions in the totality of presently-habitable near-Earth space. How can such a situation be rationalized?

The Communications Ring has a minor effect and could be tolerated as well as most LEO satellite arrays. But, the SOR must be recognized as the only option for economically putting the mass of a Shade Ring into orbit. The Shade Ring must be recognized as such a cost-effective and environmentally-acceptable way of controlling global warming that public pressure must mount to “demand” the Ring system. To prevent competitors from “sabotaging” the information and public-relations campaigns required to get people behind such an endeavor, there must be some serious “horse trading.” The fact that it would take a decade, from political and financial commitment to the first regular payload, will ease some of the competitor’s fears. It will allow them enough time to amortize their assets and to partner with the Ring System or to otherwise position themselves to profit from it. International cooperation will be required for success.

The Sling-on-a-Ring system has many other benefits to make it attractive (getting man into space, moving major polluting industries off the Earth, obtaining natural resources from the moon, Mars, and the asteroid belt, etc.). However, many other attractive opportunities have often been “killed” by a small group that has other ideas and/or fears (some people do not find Saturn’s rings attractive and will fight – on aesthetic or religious grounds - to prevent Earth from becoming so adorned). The requirement for an answer to Global Warming may be the only way for the full Ring System to be accepted without major delays and years of hard politicking.

Technical difficulties of the SOR are likely to be minor compared to those mentioned above. One potential technical problem is the mass associated with compressive members required for the SOR. For each Sling at least two 100 km "spreaders" are required (one at the base of each triangle in Fig.2) to keep the sides of the H apart. The actual compressive loads are not that great; but buckling modes must be absolutely suppressed. For a 1-stage SOR, the crosspiece of the H may also have to be semi-rigid, self-supporting, and able to withstand large lateral loading with only moderate give. While a major portion of the stability will come from tension, and therefore will benefit greatly from CCT, there are still compressive elements that will be massive simply because of their large dimensions (100km). These compressive requirements could perhaps be best treated by use of self-adjusting "tensegrity" structures [27] that are based on tension and provide both lateral and longitudinal stability with a minimal mass.

While the LEO Archipelago development is based on the high specific tensile strength of CCT, the mass of the compressive elements can be readily addressed by the low density of CCT as a replacement for glass- or carbon-fiber-reinforced plastic in tubes used as the compressive elements for the tensegrity structures. Based on densities and other properties of the CCT, it is anticipated that the mass saved in the compressive elements (without loss of strength) could be factor of 3-10 relative to fiberglass or aluminum. Thus, since the compressive members of the Sling-on-a-Ring may be its most massive component, the use of CCT can make a major contribution in this area as well.

### 5.2 System development

The Ring Systems will grow over time in several ways. Examples are given that are subject to analysis. The Communications Ring is a test bed initially. As such, it will be started in an easily accessible orbit, but one that is above most present hardware in space. Once commercially operative, it will be expanded in capability and altitude over time.

Construction in space of single-stage Sling-on-a-Ring components would follow completion of the testing and implementation of the communications ring. With an understanding and confirmation of ring dynamics, portions of the sling concept can be tested without requiring a full ring. While the sling dynamics are being tested and confirmed, the communications ring will be expanded (lengthened and raised from a low "starting" orbit into a higher one) and the other components of the initial Sling-Ring and Sling will be orbited and assembled. The initial system will not be full capacity. There may be several years of testing with small payloads and perhaps at lower-than-final orbits before all of the technology is considered to be "mature." During this Sling-testing period, the Sling-Ring will gradually be raised to full capacity (in terms of mechanical strength and electrical power) and Sling components will be introduced as they mature. The non-sling components must be distributed about the ring appropriately so that it remains stable and within orbit "bounds" as the Sling-Ring evolves.

The Sling-on-a-Ring will start with lower-orbit operation (giving time for the Comm Ring(s) to be moved to higher orbit) and with lower payloads. As other systems presently in LEO are retired and replaced by the Comm Ring(s), the SOR orbit will be raised along with the power capability, payload, and length of the sling tether. This growth might be as a result of adding new sling sub-systems (at about 3000 km each, until they start to be joined). As the sub-systems spacing decreases, the need for, and compressive strength of, spreaders is reduced and eventually the main ring will be replaced by a pair of rings.

Initial payloads (after testing is completed) could start at 1 ton each and, with experience, would rapidly grow to 10 tons and beyond. The starting aircraft (the Boeing 737 in freighter mode) has a normal payload capacity of over 20 MT [28]. With modifications for the application (high altitude, most fuel-efficient speed, and short range), this aircraft payload could probably be extended beyond 30 MT. Some aircraft would be modified to be large-volume rather than large-mass carriers. With 20 operational slings (out of ~40 on the ring), averaging 25 tons per lift, and two lifts per day, a 1000 MT per day goal for the year 2050 is reasonable. To put this in perspective, 1000 MT is only a few percent of the daily mass of air freight carried, for longer distances, by only one US company (Federal Express) [29].

As the number of slings and power systems on the ring grows, the ability to absorb larger payloads grows as well. The sling design does not change for larger payloads and with time new slings capable of handling them would be introduced. Extrapolating to the second half of the century, but with present-day aircraft (e.g., a 747 freighter, modified as above), a maximum 200 MT payload (and much larger volumes) could be expected. With these assumptions (including a higher usage factor for the slings), 10,000 MT per day into LEO could be envisioned. If the cycle time between payloads is reduced, the daily tonnage would go up accordingly.

While the lifting capability of the Sling-on-a-Ring had been explored, the return-to-earth capability has not. This certainly must be a desirable possibility. Specialized aircraft might have to be designed for that task. However, the appropriate archetype already exists in the prop-driven regime. Other goals worth studying include the quick-release option (within a cycle or two after payload pickup) to transfer payloads to higher orbits rather than simply capturing them to the Lift Ring in LEO. Another is the direct application of the Ring technology to the moon for example (where even aircraft are not required for transfer to the surface). The slings would be important for returning payloads to the Earth or pushing them on to the planets. This would certainly aid in man's access to new worlds and their raw materials.

Cost analyses and system tradeoffs must be made to determine feasibility and options. Some of this has been touched on here and in the authors' prior work. One item that had promise of providing sufficient funding for the system was the carbon credit. That may "die" before the Shade-Ring System ever gets off the ground. Some of the proposed space-solar-power technology has shown proof of concept, but no guarantees for the scalability and efficiencies required to give confidence that it could replace fossil fuels as the principle source of terrestrial electric power by the end of the century. The total system costs for a "starting" level Sling-on-a-Ring is small relative to many projects being undertaken today. But, without a guarantee of a workable system, of international agreements, of a market, and of an acceptance for the "takeover" of the LEO environment, private investment of this magnitude is unlikely.

There are some markets to be explored that could make a Shade Ring self supporting. As an example, if precessing-Slant Rings can be configured and controlled to shade major cities for 4 hours a day during the summer, how much electricity would be saved in air conditioning costs alone and how much cooler would the cities be during the summer? What would a city be willing to pay for such a service (that would greatly reduce electricity demand at peak hours)? What would be the costs associated with putting up, controlling, and servicing such a ring? A number of such cities supporting Slant Rings would have an overall effect of cooling the summers and aiding the goal of a growing Shade Ring to delay the melting of the ice sheets. While "sale" of such rings could draw material and mass-lifting resources from the world's Shade Ring, the input of independent capital could help the whole system grow more rapidly than might be possible based on just an international body's finance and control.

The opportunities seem unbounded; if the concept is not unfounded.

## 6. SUMMARY and CONCLUSIONS

This paper has presented a brief overview of a multi-purpose system that has not been here-to-fore explored as a whole. Parts of it have been explored in other contexts, and some in much greater detail, e.g. [7]. It is believed that the LEO Ring System has a synergism that overcomes the weaknesses that could impede acceptance or development of the individual rings. Not only do the different rings benefit each other, but the timing of material and political developments seem to be favoring consideration of a new approach to getting man into space.

The Comm-Ring is the most independent of the Rings. However, it provides the early experience and training with ring systems as well as early financial returns. The massive Shade and Power Rings can counter Global warming and the dependence on Earth-based fossil and nuclear fuels. However, they both require a lower-cost means of getting mass into space. On the other hand, this requirement provides the incentive (and a financial basis) for development of the Sling-on-a-Ring.

New materials have made major impacts on system feasibility, design, and costs. Starting a system based on any one technology may end up wasting years of effort if the technology does not develop. Not starting a system, because materials have not yet reached production levels, could be wasting years of profits and other benefits [5]. Major systems with critical near-term implications that require international agreement and cooperation cannot afford the luxury of waiting for a proven capability. The Sling-on-a-Ring mass lifter also has implications for the far future. But, it has the ability in the near future to meet urgent, present, needs. New materials are critical to its timely success. It would appear that a new contender, Colossal Carbon Tubes, may outpace the earlier-promising, but presently "stalled," Carbon-Nanotube-fiber technology for this application. The excellent CCT performance in a macroscopic fiber is a strong signal to proceed with detailed analysis of the LEO Archipelago system.

ACKNOWLEDGEMENT

This work is supported in part by HiPi Consulting, New Market, MD, USA, by the Science for Humanity Trust, Bangalore, India, and by the Science for Humanity Trust, Inc, Tucker, GA, USA.